\documentclass[epj,nopacs,final]{svjour}
\usepackage{graphics}
\usepackage{epstopdf}
\usepackage{latexsym,overpic,rotating}
\usepackage{subfigure,color,showkeys}
\usepackage{epsfig,color,rotating,amsmath,delarray,array,multirow}
\usepackage{makeidx,pifont,float,amssymb}

\def\ShowAnnotationsVersion{y}    
\def\AnswerYes{y}  
\ifx\ShowAnnotationsVersion\AnswerYes
 \usepackage{ulem}
  \newcommand{\delete}[1]{\sout{#1}}            
  \renewcommand{\emph}[1]{\textit{#1}}           
  \newcommand{\comment}[1]{{\color{blue}\textit{#1}}}
 \else
  \newcommand{\comment}[1]{}

  \newcommand{\delete}[1]{}
\fi

\title{Determination of the scalar polarizabilities of the proton using beam asymmetry \boldmath$\Sigma_{3}$ in Compton scattering}
\titlerunning{Measurement of the magnetic polarizability of the proton}
\authorrunning{A2 Collaboration}

\author{The A2 Collaboration \medskip \\
V.~Sokhoyan \inst{1,2},
E. J.~Downie\inst{2},
E.~Mornacchi \inst{1},
J. A.~McGovern \inst{3},
N.~Krupina \inst{1},
F.~Afzal \inst{4},
J.~Ahrens \inst{1},
C. S.~Akondi \inst{5},
J. R. M.~Annand \inst{6},
H. J.~Arends \inst{1},
R.~Beck \inst{4},
A.~Braghieri \inst{7},
W. J.~Briscoe \inst{2},
F.~Cividini \inst{1},
C.~Collicott \inst{1, 2, 8, 9},
S.~Costanza \inst{7},
A.~Denig \inst{1},
M.~Dieterle \inst{10},
M.~Ferretti \inst{1},
S.~Gardner \inst{6},
S.~Garni \inst{10},
D. I.~Glazier \inst{6, 11},
D.~Glowa \inst{11},
W.~Gradl \inst{1},
G.~Gurevich \inst{12},
D.~Hamilton \inst{6},
D.~Hornidge \inst{13},
G. M.~Huber \inst{14},
A.~K\"aser \inst{10},
V. L.~Kashevarov \inst{1},
I.~Keshelashvili \inst{10},
R.~Kondratiev \inst{12},
M.~Korolija \inst{15},
B.~Krusche \inst{10},
V.~Lensky \inst{1, 3, 16, 17},
J.~Linturi \inst{1},
V.~Lisin \inst{18},
K.~Livingston \inst{6},
I. J. D.~MacGregor \inst{6},
R. ~Macrae \inst{6},
D. M.~Manley \inst{5},
P. P.~Martel \inst{1, 13, 19},
D. G.~Middleton \inst{1, 13},
R.~Miskimen \inst{20},
A.~Mushkarenkov \inst{1, 18, 20},
A.~Neiser \inst{1},
M.~Oberle \inst{10},
H.~Ortega Spina \inst{1},
M.~Ostrick \inst{1},
P.~Ott \inst{1},
P. B.~Otte \inst{1},
B.~Oussena \inst{1, 2},
D.~Paudyal \inst{14},
P.~Pedroni \inst{7},
A.~Polonski \inst{12},
S.~Prakhov \inst{1, 2, 21},
A.~Rajabi \inst{20},
T.~Rostomyan \inst{10},
A.~Sarty \inst{9},
S.~Schumann \inst{1},
K.~Spieker\inst{4},
O.~Steffen \inst{1},
I. I.~Strakovsky \inst{2},
B. ~Strandberg \inst{6},
T.~Strub \inst{10},
I.~Supek \inst{15},
A.~Thiel \inst{4},
M.~Thiel \inst{1},
A.~Thomas \inst{1},
M.~Unverzagt \inst{1},
S.~Wagner \inst{1},
D. P.~Watts \inst{11},
J.~Wettig \inst{1},
L.~Witthauer \inst{10},
D.~Werthm\"{u}ller \inst{10, 6},
M.~Wolfes \inst{1}, and
L.~Zana \inst{1}
         \\
         }
\institute{\inst{1} Institut f\"{u}r Kernphysik, Universit\"{a}t Mainz, D-55099 Mainz, Germany\\
\inst{2} Department of Physics, The George Washington University, Washington, DC 20052, USA\\
\inst{3} School of Physics and Astronomy, University of Manchester, Manchester, M13 9PL, UK\\
\inst{4} Helmholtz-Institut f\"{u}r Strahlen- und Kernphysik, Universit\"{a}t Bonn, D-53115 Bonn, Germany\\
\inst{5} Department of Physics, Kent State University, Kent, Ohio 44242, USA\\
\inst{6} SUPA School of Physics and Astronomy, University of Glasgow, Glasgow G12 8QQ, UK\\
\inst{7} INFN Sezione di Pavia, I-27100 Pavia, Italy\\
\inst{8} Department of Physics and Atmospheric Science, Dalhousie University, Halifax, Nova Scotia B3H 4R2, Canada\\
\inst{9} Department of Astronomy and Physics, Saint Mary's University, Halifax, Nova Scotia B3H 3C3, Canada\\
\inst{10} Departement Physik, Universit\"{a}t Basel, CH-4056 Basel, Switzerland\\
\inst{11} School of Physics, University of Edinburgh, Edinburgh EH9 3JZ, UK\\
\inst{12} Institute for Nuclear Research, 125047 Moscow, Russia\\
\inst{13} Department of Physics, Mount Allison University, Sackville, New Brunswick E4L 1E6, Canada\\
\inst{14} Department of Physics, University of Regina, Regina, Saskatchewan S4S 0A2, Canada\\
\inst{15} Rudjer Boskovic Institute, HR-10000 Zagreb, Croatia\\
\inst{16} Institute for Theoretical and Experimental Physics, 117218 Moscow, Russia\\
\inst{17} National Research Nuclear University MEPhI (Moscow Engineering Physics Institute), 115409 Moscow, Russia\\
\inst{18} Lebedev Physical Institute, 119991 Moscow, Russia\\
\inst{19} Laboratory for Nuclear Science, Massachusetts Institute of Technology, Cambridge, Massachusetts 02139, USA\\
\inst{20} Department of Physics, University of Massachusetts Amherst, Amherst, Massachusetts 01003, USA\\
\inst{21} Department of Physics and Astronomy, University of California Los Angeles, Los Angeles, California 90095-1547, USA
}

\date{Received: date \today / Revised version: date}
\abstract{
The scalar dipole polarizabilities, $\alpha_{E1}$ and $\beta_{M1}$, are fundamental properties
related to the internal dynamics of the nucleon. The currently accepted values of the 
proton polarizabilities were determined by fitting to unpolarized proton 
Compton scattering cross section data. 
The measurement of the beam asymmetry $\Sigma_{3}$ in a certain
kinematical range provides an alternative approach to the extraction of the scalar polarizabilities.
At the Mainz Microtron (MAMI) the beam asymmetry was measured for Compton scattering 
below pion photoproduction threshold for the first time. The results are compared with model calculations and the influence
of the experimental data on the extraction of the scalar polarizabilities is determined. 
\PACS{{PACS-key}{describing text of that key}   \and
      {PACS-key}{describing text of that key}
     } 
} 
\begin{document}
\maketitle
\section{Introduction}
\label{intro}

The proton polarizabilities characterize the inelastic structure of the nucleon \cite{Bernard:1995dp,{Hagelstein}}, and thus are
complementary to the electromagnetic form factors that describe the elastic structure.
These polarizabilities are fundamental properties of the proton, as much as its mass, charge, and magnetic moment.
They contribute at second order in the low-energy expansion of Compton scattering on the proton ($\gamma p \to \gamma p$) \cite{Holstein1,{Babusci}},
and are the first contributions beyond the low-energy theorem \cite{Low,{GellMann_Goldberger}}. 
The scalar polarizabilities are of great importance for nuclear and atomic physics, and other related fields, 
and are currently a significant source of uncertainty in the determination of the proton charge radius from muonic hydrogen Lamb shift \cite{Antognini}. 
The neutron polarizabilities even play a role in neutron star physics (see e.g. \cite{Bernabeu}), 
illustrating the impact of these sub-microscopic properties at macroscopic scales. Furthermore, a precise and model-independent determination of the scalar
polarizabilities is crucial for the extraction of the spin polarizabilities. The first measurement of the double polarization asymmetry
$\Sigma_{2x}$ at MAMI and the individual extraction of the proton spin polarizabilities was reported recently \cite{PMartelSpin}. 
The current values of the scalar polarizabilities of the proton, presented by the Particle Data Group (PDG) \cite{PDGCite2014}, are
\begin{eqnarray*}
\alpha_{E1} = (11.2 \pm 0.4)\times 10^{-4} \rm \, fm^{3} \\
\beta_{M1} = (2.5 \pm 0.4)\times 10^{-4} \rm \, fm^{3}.
\label{Eq: alpha beta}
\end{eqnarray*}
These values were extracted from the unpolarized cross section of Compton scattering on the
proton at energies below 170 MeV, from a large number of experiments, but particularly those of 
Refs.~\cite{OlmosCross,{FederspielCross},{MacGibbonCross},{ZiegerCross}}. 
By far the largest of these data sets \cite{OlmosCross} was measured previously at MAMI.

The measurement of the beam asymmetry $\Sigma_{3}$ provides an alternative approach to the extraction of the
scalar polarizabilities, with a potentially complementary sensitivity to $\alpha_{E1}$ and $\beta_{M1}$. A low-energy expansion 
of $\Sigma_{3}$ gives \cite{KrupinaSigma3} 
\begin{equation}
    \Sigma_{3} = \Sigma_{3}^{\rm (B)} - \frac{4M\omega^{2}\cos\theta\sin^{2}\theta}{\alpha_{em}(1 + \cos^{2}\theta)^2}\beta_{M1} + {\cal O}(\omega^{4}),
      \label{Eq: Sigma3 expansion}
\end{equation}
where $\Sigma_{3}^{\rm (B)}$ represents the Born contribution, {\it M} is the mass of the target, $\alpha_{em}$ is the fine-structure constant, 
and $\omega$ and $\theta$ are the incoming photon energy and the polar angle of the outgoing photon in the Breit frame. The Born term $\Sigma_{3}^{\rm (B)}$ 
depends only on the mass, charge, and magnetic moment of the proton. At sufficiently low energies ($\omega \ll m_{\pi}$), the contribution of the terms 
at ${\cal O}(\omega^{4})$ could be neglected and very precise measurements could isolate $\beta_{M1}$ cleanly. 
In practice, even though $\alpha_{E1}$ and other terms (such as the pion-pole contribution, in which both photons couple to the nucleon 
via the exchange of a neutral pion) only enter at ${\cal O}(\omega^4)$, for energies above about 80~MeV they are at least as 
important as $\beta_{M1}$ \cite{McGovernEPJ}. However, in the context of a theoretical prediction for the asymmetry, $\alpha_{E1}$ and $\beta_{M1}$ can both be fit. 
In this work, the first results on the beam asymmetry below pion photoproduction threshold are presented and compared 
with existing theoretical predictions. The influence of the new data
on the polarizabilities and the potential of further high-precision measurements are discussed.

\section{Experiment}
The experiment was performed at the MAMI accelerator facility \cite{KaiserMAMI,{JankowiakMAMI}} using an electron beam
with an energy of 883~MeV.  The electron beam passed through a thin diamond radiator, 
where some of the electrons underwent coherent bremsstrahlung  \cite{TimmPol,{LohmannPol}}. The energy-degraded electrons were then detected in the focal plane of the 
Glasgow Mainz Photon Tagging Spectrometer \cite{AnthonyTagg,{HallTagg},{McGeorgeTagg}}. 
Using energy conservation, the energy of the resulting photon is determined.  The linearly polarized photons traveled 
down the beamline to the experimental area, where they impinged on a 10-cm-long liquid 
hydrogen target. The photons were then energy-tagged to the reaction products using a
timing coincidence between the tagger and the detector system. 

\begin{figure}[hbtp]
\begin{center}
\includegraphics[width=\columnwidth]{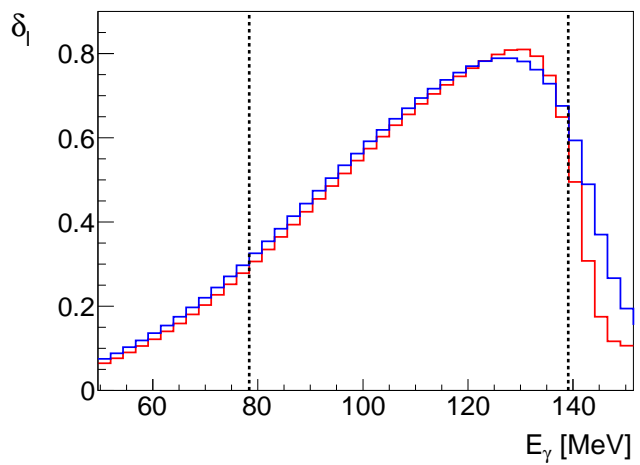}
\end{center}
\caption{Degree of linear polarization for two orientations of the diamond with polarization plane parallel (red line) and perpendicular 
(blue line) to the horizontal lab axis. The black dashed lines indicate the range used in further analysis.}
\label{polarization}
\end{figure}

The measurements were performed with two orientations of the diamond, resulting in a relative angle 
of $90^{\circ}$ between the corresponding polarization planes (formed by the momentum of the incoming photon and its electric field vector).
The degree of linear polarization was determined directly from the experimental data, by fitting the polarization enhancement obtained from the ratio of the 
photon energy spectra taken with the diamond and a reference copper radiator. The enhancement was calculated  in short time intervals corresponding to the readout 
of the scaler counters (typically every 1-2 seconds).
This event-based method allowed us to account for changes in the values of linear polarization due to small variations in the position of the incoming electron beam. 
Figure \ref{polarization} shows the resulting averaged degree of linear polarization for both polarization settings. These values were used in the determination of
the beam asymmetry (see Sec.~\ref{sect: beam asymm}). 
The reaction products were detected by the Crystal Ball/TAPS calorimeter system covering $97 \%$ of the full solid angle.
The Crystal Ball calorimeter, consisting of 672 NaI crystals, covered polar angles from $20^{\circ}$ to $155^{\circ}$ \cite{StarostinCB}.
The TAPS forward wall is built of 366 $\rm BaF_{2}$ crystals covering polar angles from $4^{\circ}$ to $20^{\circ}$, and 
72 $\rm PbWO_{4}$ crystals covering the angular range from $1^{\circ}$ to $4^{\circ}$ \cite{NovotnyTAPS}. 
In this work, we used the signal from the Crystal Ball and the $\rm BaF_{2}$ crystals. Both Crystal Ball and TAPS detectors have full azimuthal coverage. 
In addition to these calorimeters, a Particle Identification Detector (PID), consisting of 24 scintillator bars \cite{WattsPID}, and two Multiwire Proportional 
Chambers (MWPCs), were incorporated in the Crystal Ball detector system for identification and enhanced tracking of charged particles. 
In the TAPS region, the charged particles were identified with thin plastic veto scintillators placed in front of the $\rm BaF_{2}$ crystals.

In the selection of $\gamma p \to \gamma p$ events, precisely one neutral hit in the Crystal Ball was required since the outgoing proton did not
reach the calorimeters in the relevant energy range ($E_{\gamma}$ = 79--139 MeV). For the identification of the photon, it was required that no
charged hit was identified in the PID or MWPCs.
Due to significant contamination of the forward region with electromagnetic background originating from the photon beam, only events with outgoing
photon scattering angles $\theta > 30^{\circ}$ were used in the extraction of the beam asymmetry. In order to reduce the random background
in the Crystal Ball, it was required that the scattered photon was time-coincident with the tagger hit within 3 ns. In order to remove randomly coincident events
from the selection, we sampled the random background in two timing windows at (-200; -20) ns and (20; 200) ns and subtracted
it from the signal after normalization according to the width of the selected time intervals. The data were divided into three incoming photon energy ranges
($E_{\gamma}$ = 79--98 MeV, 98--119 MeV, and, 119--139 MeV). For each of these ranges, the energy of the outgoing photons $E_{\gamma}^{\text {out}}$ 
had to agree with limits determined from Monte Carlo simulations ($E_{\gamma}^{\text {out}}$ = 65--100 MeV, 80--120 MeV, and 95--140 MeV respectively).
These limits were defined according to the kinematics of Compton scattering and energy resolution of the Crystal Ball. Furthermore, in order to remove events from 
the kapton windows of the target vessel and scattering chamber, (positioned along the photon beam), the same analysis procedure was also applied to data taken with 
an empty target and the contribution of the empty target was scaled and subtracted.

\begin{figure}[pt]
\begin{center}
\includegraphics[width=\columnwidth]{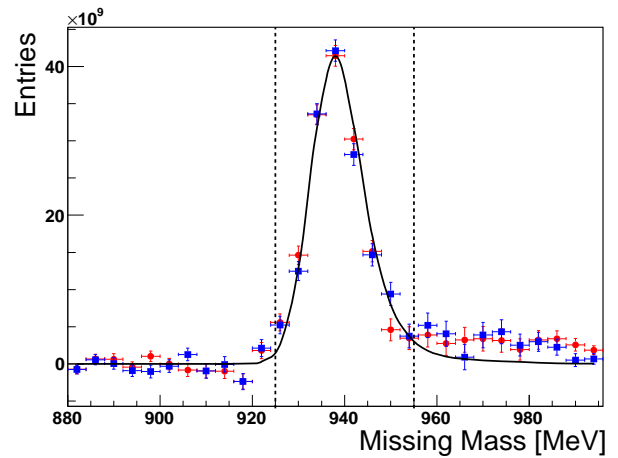}
\end{center}
\caption{Missing mass for the incoming photon energy range 79--98 MeV and scattered photon polar angles
$30^{\circ} < \theta < 155^{\circ}$, data were obtained for two polarization planes 
parallel and perpendicular to the horizontal (red circles and blue squares respectively). 
The black curve represents the Monte Carlo simulated distribution for Compton scattering, the dashed lines indicate the applied cut.}
\label{missing mass}
\end{figure}

The consideration of the missing mass spectrum before the outgoing photon
energy cut, as shown in Fig.~\ref{missing mass}, allowed us to determine the 
upper limit of the background contamination in the data (after the application 
of this cut, the tails in the missing mass spectrum are suppressed, but the signal region remains practically unchanged).
The data obtained with each of the two polarization planes (data were also taken with an unpolarized photon beam from a copper radiator)
are compared with the spectrum obtained from Monte Carlo simulation of Compton scattering.
The agreement of the lineshape between both data sets indicates an absence of strong systematic effects in the unpolarized component of the
selected data sample.  For all three photon energy ranges used to determine $\Sigma_{3}$, a missing mass cut of 
$925 < M_{miss} < 955$ MeV was applied (see Fig.~\ref{missing mass}). The shapes of the distributions obtained from the experimental data 
and Monte Carlo simulation are in good agreement, indicating no significant background contamination in the selected range.
The overall background in the final data sample, consisting of 200,143 
Compton scattering events in the energy range ($79 < E_{\gamma} < 139$ MeV), with polar angle coverage {($\rm 30^{\circ} < \theta < 155 ^{\circ}$)}, did not exceed $4\%$. 
We accounted for the possible influence of the remaining background on the final results in the systematic error (see Sec.~\ref{sect: beam asymm}).
Furthermore, in order to test for systematic effects in the identification of Compton 
scattering events, data were also taken with unpolarized beam from a copper radiator. 
From these data, the unpolarized cross sections were extracted. The measured unpolarized cross sections 
agreed well with previous data \cite{OlmosCross}.

\begin{figure}[pt]
\begin{center}
\includegraphics[width=\columnwidth]{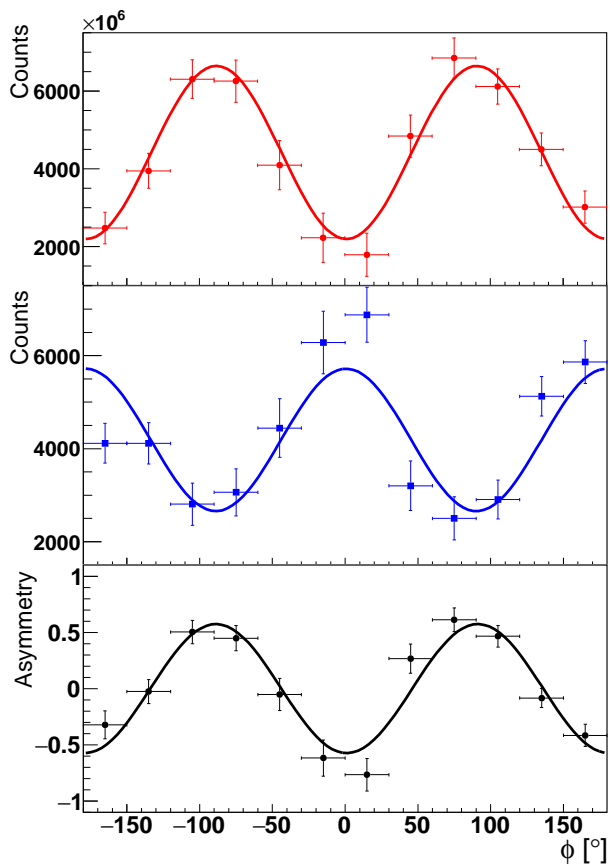}
\end{center}
\caption{Sample $\phi$ distributions of the outgoing photon. The upper two panels show $\phi$ distributions
obtained with polarization plane parallel (upper) and perpendicular to (middle) the horizontal lab axis. 
The lower-most panel shows the $\phi$-dependent asymmetry obtained from Eq.~\ref{Eq: Sigma3 determine}.}
\label{Fig: phi}
\end{figure}

\section{Beam asymmetry \boldmath$\Sigma_{3}$}
\label{sect: beam asymm}
The use of a linearly polarized photon beam with an unpolarized target introduces an azimuthal angle 
dependence to the cross section.  We are interested in  the case of two perpendicular polarization 
directions, corresponding to perpendicular orientations of the diamond crystal. Defining $\phi$ as 
the angle between one of the polarization planes and the scattering plane defined by the incoming and
outgoing photon momenta, the two cross sections can be written as 
\begin{equation}
 \sigma_{pol} = \sigma_{unpol}(1 \pm \delta_{\,l}\, \Sigma_{3} \, {\cos}\, 2\phi),
\label{Eq: Pol Cross}
\end{equation}
where $ \sigma_{unpol}$ is the cross section for an unpolarized photon beam and $\delta_{\,l}$ is the degree of linear polarization of the beam.
The beam asymmetry $\Sigma_{3}$ determines the magnitude of the modulation.
The top two panels of Fig.~ \ref{Fig: phi} show sample $\phi$ distributions, in which the ${\cos}\, 2\phi$ modulation can be clearly identified.

At $\phi = 0^{\circ}$ the photon polarization is either parallel or perpendicular to the scattering plane, with corresponding cross sections denoted 
$\sigma_{\parallel}$ and $\sigma_{\perp}$, giving the usual expression for $\Sigma_{3}$: 
\begin{equation}
 \Sigma_{3} \equiv \frac{\sigma_{\parallel} - \sigma_{\perp}}{\sigma_{\parallel} + \sigma_{\perp}}.%
\label{Eq: Sigma3 def}
\end{equation}

In order to account for different polarization values for the sets corresponding to the two different orientations of the diamond, the corresponding number of
Compton scattering events was weighted with the corresponding linear polarization values (where the $\Sigma_{3}\cos{2\phi}$ term follows
from Eq.~\ref{Eq: Pol Cross}):
\begin{equation}
\begin{split}
 \Sigma_{3}\cos(2(\phi + \phi_{0})) =\frac{\sigma_{\parallel} - \sigma_{\perp}}{\delta_{\perp}\sigma_{\parallel} + \delta_{\parallel}\sigma_{\perp}} =
 \frac{N_{\parallel} - N_{\perp}}{\delta_{\perp}N_{\parallel} + \delta_{\parallel}N_{\perp}},
\end{split}
\label{Eq: Sigma3 determine}
\end{equation}
here  $\delta_{\parallel}$ and $\delta_{\perp}$ represent the degree of polarization for the two polarization settings respectively, and
$\rm N_{\parallel}$ and $\rm N_{\perp}$ represent the event rates for the two polarization settings measured in the experiment. 
The effect of a possible deviation of the polarization planes from the expected positions (e.g. due to geometrical positioning of the radiator) 
was accounted for by fixing the phase of the azimuthal angle to the value determined by fitting the experimental data ($\phi_{0} = (-1 \pm 1)^\circ$).
The bottom panel of Fig.~ \ref{Fig: phi} shows the angle-dependent asymmetry obtained according to Eq. \ref{Eq: Sigma3 determine}; the ${\cos}\, 2\phi$ 
modulation is again clearly seen.

\begin{figure}[pt]
\begin{center}
\includegraphics[width=\columnwidth]{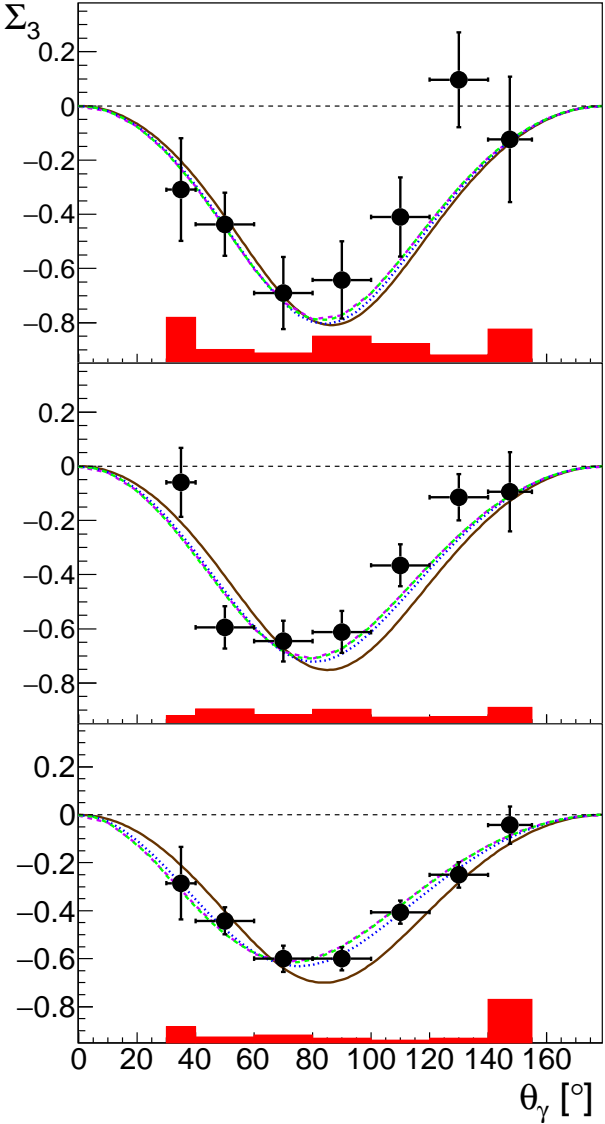}
\end{center}
\caption{Beam asymmetry $\rm \Sigma_{3}$ for three energy ranges (uppermost: 79--98 MeV, middle: 98--119 MeV, lowermost: 119--139 MeV). The errors represent
statistical errors, the red bars indicate the systematic error. Green dashed curve: BChPT calculation  \cite{Lensky:2009uv},
magenta dashed-dotted: DR calculation \cite{DrechselPol,{PasquiniPol}},  blue dotted: HBChPT  \cite{McGovernEPJ}, 
all with $\alpha_{E1} = 10.65 \times 10^{-4} \rm \, fm^{3}$ 
and $\beta_{M1} = 3.15 \times 10^{-4} \rm \, fm^{3}$; brown solid: Born term (curves correspond to the central values of the shown energy bins).}
\label{fig: Sigma3}
\end{figure}

The beam asymmetry was extracted by fitting the $\phi$ distributions and equating to $(3/\pi)\Sigma_{3}\cos(2(\phi + \phi_{0}))$ in Eq.~\ref{Eq: Sigma3 determine}. 
The prefactor of $\sin(\Delta\phi)/\Delta\phi=(3/\pi)$ accounts for the damping of the amplitude of the modulation from averaging over bins of width 
$\Delta\phi=30^\circ$.  The results are shown in Fig.~\ref{fig: Sigma3} for various ranges of the incoming photon energy and polar angle of the scattered 
photon. The error bars shown are statistical; the systematic errors are presented as red bars.
The most significant contribution to the systematic error is due to the normalization of the photon flux (typically in order of 15\% at low energies 
and 4\% at higher energies). The next most significant source of uncertainty is background contamination, which was estimated based on the events outside 
(on the right) of the missing mass peak (increasing up to 50\% at low and high polar angles and being in order of 3\% for central angular bins). 
Another source of uncertainty is the determination of the degree of polarization. Generally, the influence of this effect is small (typically below 2\%)
compared to the other effects. The combination of the effects described above results in the systematic error shown in Fig.~\ref{fig: Sigma3}.

In addition, Fig.~\ref{fig: Sigma3} also shows a comparison between the data and various calculations.
The solid line shows only the Born term (see Eq. ~\ref{Eq: Sigma3 expansion}), which is independent of all proton structure except the anomalous magnetic moment.
Fig.~\ref{fig: Sigma3} also shows predictions for the asymmetry from Dispersion Relations (DR, dotted) \cite{DrechselPol,{PasquiniPol}} and from two variants of 
Chiral Perturbation Theory:  Baryon Chiral Perturbation Theory (BChPT, dashed) \cite{Lensky:2009uv} and 
Heavy Baryon Chiral Perturbation Theory (HBChPT, dash-dotted) \cite{McGovernEPJ}. 
In the former variant the nucleon is treated relativistically whereas in the latter an expansion of the amplitudes in powers of the inverse nucleon 
mass is performed; in addition the latter calculation is carried out to one order higher in their common power-counting than the former.
Both include the contribution of the Delta isobar. In each of these calculations the spin polarizabilities are not free parameters, 
but predictions of the theory (their values can be found in \cite{McGovernEPJ,{Lensky2015},{PasquiniPol}} and in particular are summarized in table IV of \cite{Lensky2015}). 
However a check within the BChPT framework showed that at the present level of accuracy the choice of the spin polarizabilities has no noticeable influence on the fit 
to the data of Fig.~\ref{fig: Sigma3}.

\begin{figure}[hbtp]
\begin{center}
\includegraphics[width=\columnwidth]{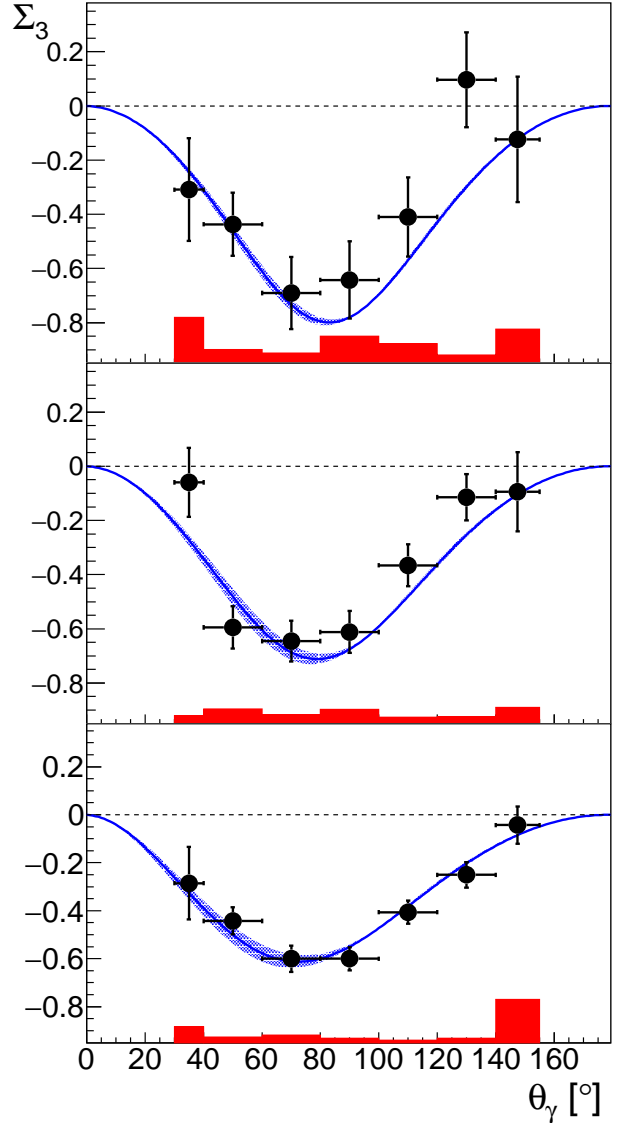}
\end{center}
\caption{The result of the fit within BChPT framework obtained by averaging the numerator and denominator in Eq.~\ref{Eq: Sigma3 determine}
over angle and energy (blue curve). Shaded bands are determined by the error in $\beta_{M1}.$  Notation for the data as in Fig.~\ref{fig: Sigma3}.}
\label{fig: Sigma3_fit}
\end{figure}

The data are in  good agreement with the results obtained in all theories. At the same time, the clear separation between the data and the
contribution of the Born term, particularly in the energy range  $E_{\gamma}$ = 119--139 MeV, indicates the sensitivity of the
data to dynamical effects including but not limited to the scalar polarizabilities.

Rather than taking the polarizability values from elsewhere, they can be fit directly to the asymmetry data; however the current precision means that such an 
extraction will not be competitive.  A two-parameter fit gives values of both $\alpha_{E1}$ and $\beta_{M1}$, but to minimize the uncertainty we prefer to use 
the well-established value  $\alpha_{E1}+\beta_{M1}=14.0\times 10^{-4} \rm \, fm^{3}$ obtained from the Baldin sum rule \cite{BaldinSumRule} and perform a 
one-parameter fit, giving a value for $\alpha_{E1}-\beta_{M1}$ or, equivalently, $\beta_{M1}$.

The fit was performed using only the new data on the beam asymmetry within the BChPT \cite{Lensky:2009uv,{Lensky2015},{Lensky:2014efa}} 
and HBChPT frameworks \cite{McGovernEPJ} (combining the statistical and the systematic errors quadratically), with respective results 
$\beta_{M1} \rm = 2.8\substack{+2.3 \\ -2.1}\times 10^{-4}\, fm^{3}$ ($\chi^{2}/ndf = 19.2/20$) and  $\beta_{M1} \rm = 3.7\substack{+2.5 \\ -2.3}\times 10^{-4}\, fm^{3}$
($\chi^{2}/ndf = 17.1/20$); the error is obtained from the $\chi^{2}_{min} + 1$ interval. Because of the finite bin widths, theoretical predictions for the asymmetries, 
which are ratios of the difference and sum of the two polarized differential cross sections (see Eq.~\ref{Eq: Sigma3 determine}), have been obtained by averaging 
the numerator and denominator over angle (weighted by $\sin\theta$)  and energy. Because the asymmetry varies quite rapidly with angle, this makes a noticeable difference 
to the results. 

Figure \ref{fig: Sigma3_fit} shows the BChPT result, and  the sensitivity of the beam asymmetry to the magnetic
polarizability $\beta_{M1}$ is indicated by the width of the shaded band.  The corresponding plot for the HBChPT fit is very similar.

The results for $\beta_{M1}$ obtained within both frameworks are compatible with each other and also agree with the 
$\beta_{M1} \rm = (2.5 \, \pm \, 0.4) \times 10^{-4} \, fm^{3}$ provided by the PDG \cite{PDGCite2014}. Despite the fact that presently the  large
errors mean that our determination is not competitive and does not significantly impact a global determination, the results indicate that the observable $\Sigma_{3}$ 
provides new input for the determination of the magnetic polarizability $\beta_{M1}$. In order to achieve high accuracy, new high-precision measurements are foreseen 
at MAMI both with significantly higher statistics and with improved control over systematic effects (e.g. due to the photon flux normalization and very stable linear 
polarization). The new measurement will follow the ongoing upgrade of the Crystal Ball/TAPS setup at MAMI and the 
beam asymmetry and unpolarized cross sections will be determined with unprecedented precision.

\section{Summary}

We have reported on the first measurement of the beam asymmetry $\Sigma_{3}$
for  Compton scattering below pion production threshold. The results confirm the existing theoretical predictions
 (ChPT, HBChPT, DR), and deviate notably from the Born term in which the
contributions of the polarizabilities are not included. The  results obtained show that the extraction of the
scalar polarizabilities from the beam asymmetry as an alternative to the extraction from the
unpolarized cross sections is possible, and challenge us to obtain higher-statistics data sets for this observable.
In the future, new high-precision measurements both for the beam asymmetry and unpolarized
cross section will be performed at MAMI.\\

We wish to acknowledge the outstanding support of the
accelerator group and operators of MAMI. We also acknowledge the contributions and suggestions
of H. Grie{\ss}-hammer, V. Pascalutsa, and B. Pasquini to this work.
We thank the students of Mount Allison University and The George Washington
University for their contributions to the experiment. We acknowledge support from the Collaborative Research Center (CRC) 1044
and Schweizerischer Nationalfonds. The research leading to these results has received funding from the European Community's Seventh Framework Programme FP7/2007-2013 under Grant Agreement Nr. 227431. 
V. Sokhoyan acknowledges the support of the Carl-Zeiss-Stiftung. We also also acknowledge support from the
Natural Sciences and Engineering Research Council of Canada (NSERC) FRN: SAPPJ-2015-00023.
This work was also supported by the UK Science and Technology
Facilities Council (ST/J00175/1, ST/G008604/1, ST/G008582/1, ST/J00006X/1).
This material is based upon work supported by the 
U.S. National Science Foundation under Grant Numbers PHY-1309130, IIA-1358175, 
and the U.S. Department of Energy (Offices of Science and Nuclear Physics, 
Award Nos. DE-FG02-99-ER41110, DE-FG02-88ER40415, DE-FG02-01-ER41194).

%
%

%
%
%

%
%

\end{document}